\documentclass[final]{svjour3}
\usepackage[T1]{fontenc}
\usepackage[latin1]{inputenc}
\usepackage{graphicx}
\usepackage{amssymb}
\usepackage{mathptmx}
\usepackage[numbers]{natbib}

\makeatletter
\journalname{JLTP}

\bibpunct{}{}{,}{s}{}{,}
\begin{document}

\title{Thermal compression of atomic hydrogen on helium surface}

\author{J. J\"arvinen \and  J. Ahokas \and S. Vasiliev}
\institute{Wihuri physical laboratory, Department of Physics, University of
Turku, 20014 Turku, Finland\\ 
\email{jaanja@utu.fi}}

\date{24.10.2006}

\maketitle

\begin{abstract}
We describe experiments with spin-polarized atomic hydrogen gas
adsorbed on liquid $^{4}$He surface. The surface gas density is
increased locally by thermal compression up to $5.5\times10^{12}$
cm$^{-2}$ at 110 mK. This corresponds to the onset of quantum
degeneracy with the thermal de-Broglie wavelength being 1.5 times
larger than the mean interatomic spacing. The atoms were detected
directly with a 129 GHz electron-spin resonance
spectrometer probing both the surface and the bulk gas. This, and the
simultaneous measurement of the recombination power,  allowed us to
make accurate studies of the adsorption isotherm and the heat removal
from the adsorbed hydrogen gas. From the data, we estimate the thermal
contact between 2D hydrogen gas and phonons of the helium film. We
analyze the limitations of the thermal compression method and the 
possibility to reach the superfluid transition in 2D hydrogen gas.

\end{abstract}

\section{Introduction}
Spin-polarized atomic hydrogen (H$\downarrow$) adsorbed on the surface
of superfluid helium represents an 
almost ideal realization of a two-dimensional (2D) Bose
gas. \cite{Walraven1990} The superfluid helium surface is a perfect
substrate for the 2D gas, since
it is translationally invariant and the interaction potential supports
only one bound
state for hydrogen atom. The main interest in studies of
two-dimensional Boson gas is the formation of Bose-Einstein
condensate, which is not fully understood in 2D. In contrast to
a 3D condensate, the phase of a 2D condensate fluctuates on a
macroscopic scale and there is no long range phase correlation in
the system, while the density fluctuations are fully suppressed.
The formation of superfluidity in 2D is explained by the topological
theory of Berezinskii, Kozterlitz and
Thouless\cite{Berezinskii,KK} (BKT), and recently the BKT-type
crossover has been observed in optically trapped atomic
gas.\cite{NatureBKT2006} Beside the Bose-Einstein condensation, reduced dimensionality
can be studied e.g. in the behavior of the recombination rate
constants,\cite{TurkuBKT,Khawaja2002,Prof2002} and in the strength 
of interatomic interactions.\cite{BECQ2DGas}

Reaching quantum degeneracy in H$\downarrow$ is limited by the
large energy release in atomic recombination. To remove this
obstacle, Kagan\cite{Kagan1985} \emph{et al.} proposed to use an open system,
where most of the recombination energy is carried off by the excited
H$_{2}$ molecules and the atoms escaping from the sample. In
the case of adsorbed H$\downarrow$ gas, only a small fraction of the energy
$\sim1\%$ remains on the surface to be removed
via the excitations of the helium film, ripplons and phonons. 
This has been utilized in two local compression methods: thermal and
magnetic compression. In the magnetic compression method
used e.g. by Safonov\cite{TurkuBKT} \emph{et al.} the atoms are
compressed in a small region of the sample cell (SC) wall by strong
magnetic field gradients. The sample on such a ``magnetic spot''
is at higher temperature than the rest of the cell wall, and the
recombination heat is transferred by the ripplons along the helium
surface. The cooling efficiency depends on the length of the heat
path and therefore the size of the sample was reduced to about 20 $\mu$m.
Due to the small size and the strong magnetic field gradient it
is impossible to implement sensitive spectroscopic diagnostics of
the surface atoms. In the thermal compression, first utilized by
Matsubara\cite{ColdSpot} \emph{et al.,} a small ``cold spot''
(CS) is cooled to a lower temperature than the rest of the sample
cell. The compression is caused by a reduced desorption rate of
the atoms from the CS, which decreases exponentially with
the temperature. The heat of the recombination is removed via the contact
between ripplons and phonons of the helium film. A more detailed
description of the compression methods can be found from a review
of Vasiliev and Jaakkola.\cite{Vasilyev2004}

In this paper we describe our experiments utilizing thermal
compression to increase the density of 2D atomic hydrogen gas.
General features of 2D
atomic hydrogen and the compression method are found in section 2. An
adsorption isotherm of 
H$\downarrow$ on helium surface and a ripplon-phonon thermal
contact are introduced. Section 3 contains a description of the
experimental setup where we emphasize the differences
between the two configurations of the sample cell used in the experiments. In
section 4, the measurement procedure is explained. We describe in
detail how the data are acquired and processed. In sections 5 and
6, we present the results of experiments on 2D H$\downarrow$ and
discuss the efficiency and limitations of thermal compression
to reach the BKT transition.

\section{Method of thermal compression}

\begin{figure}

\includegraphics[%
  clip,width=0.80\linewidth,
  keepaspectratio]{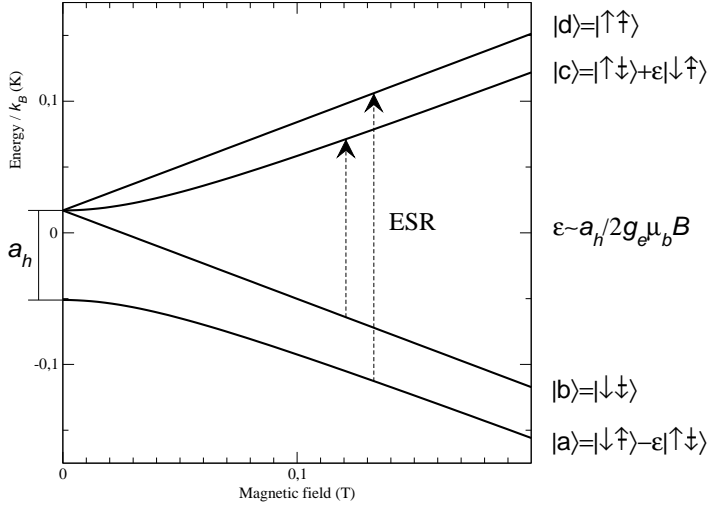}

\caption{The hyperfine energy levels and the state vectors of 
  hydrogen atom in magnetic field. The arrows show the allowed ESR
  transitions between 
  the levels.}
\label{hyper}
\end{figure}

Atomic hydrogen gas is metastable against recombination into
molecules. Formation of a hydrogen molecule in a collision always
needs at least three participating bodies. At low bulk densities
($\lesssim10^{15}$ cm$^{-3}$) considered the three particle
collisions in the bulk are fairly rare and, therefore, the recombination takes
place only in the adsorbed 2D gas phase, where the helium surface
is the required third body. On the surface the recombination proceeds
through the collisions of two or three hydrogen 
atoms. The atoms populate the lower two hyperfine states in high
magnetic field and at low temperatures. These high 
field seeking hyperfine 
states are usually denoted by letters $a$ (F=0,
m$_{\textrm{f}}$=0) and $b$ (F=1, m$_{\textrm{f}}$=-1). The hyperfine
energy levels are plotted in fig. \ref{hyper}. The state
$a$ has a small admixture of the spin state with opposite orientation of the
electron spin. This provides a finite recombination probability in
the collisions involving $a$-state atoms and leads to a reduction of the
$a$-state population to a negligible level. Then the gas becomes doubly spin
polarized (H\makebox[1.025ex][l]{$\downdownarrows$}-, where \makebox[0.25ex][l]{$\downarrow$}- marks
the nuclear spin) containing atoms only in the $b$-state, which is a pure spin
state. The atoms in this state can not form molecules in binary
collisions, thus the
decay of a doubly polarized sample is controlled by the one-body
relaxation from $b$ to $a$-state and by the dipolar three-body
recombination. The heat generated by the three-body
surface recombination increases rapidly as a third power of density, and
it is one of the major obstacles for reaching the BKT transition in
H\makebox[1.025ex][l]{$\downdownarrows$}-.

An adsorption equation, e.g. a relation between the
surface and bulk gas density in thermal equilibrium is found by equating
the chemical potentials of the gas phases.\cite{Svistunov91} Taking into account
a possible temperature difference between the surface and bulk gas
the surface density is given by\begin{equation}
\sigma=\left(\frac{T_{c}}{T_{s}}\right)^{3/2}n\Lambda\exp\left(\frac{E_{a}}{k_{B}T_{s}}\right),\label{eq:isotherm}\end{equation}
where $E_{a}$ is the adsorption energy (for H on $^{4}$He surface
$E_{a}=1.14(1)\times k_{B}$ K)\cite{Ea}, $n$ is the bulk gas
density, $T_{c}$ $(T_{s})$ is the bulk (surface) gas temperature,
and $\Lambda$ is the thermal de Broglie wavelength of the adsorbed
gas. This equation is valid in non-degenerate case
$\sigma\Lambda^{2}\ll1$. At higher degeneracy one has to use the
full quantum isotherm taking into account Bose-statistics and
interactions between the atoms.\cite{Svistunov91} The prefactor
$(T_{c}/T_{s})^{3/2}$ takes into account the temperature
difference between the bulk and surface gases. Exponential growth
of $\sigma$, proportional to $E_{a}/k_{B}T_{s}$, is essential for
the thermal compression method. Surface density is compressed by a
factor of about $100$ when the temperature is lowered,
e.g., from 100 mK to 70 mK. Eq. (\ref{eq:isotherm}) also yields
$T_{s}$ provided $E_{a}$, $\sigma$ and, $n$ are known. The
adsorption energy is known rather accurately from the
experiments\cite{Ea,VasilievPRA2004} and it is not affected by
the solid substrate beneath the liquid $^{4}$He layer unless the
latter is thinner than about 20 monolayers.\cite{Godfried1985}

About 99\% of the energy generated in the surface
recombination appear in the form of highly excited H$_{2}^{*}$
molecules, which desorb from the surface and leave the
recombination site. The energy is then gradually released in
numerous collisions with the sample cell
walls.\cite{ColdSpot,vasiliev1993,Meyer1994} Idea of
local compression is to have a large sample cell area with small
surface density through which the recombination heat is removed
and only a small high density region where the recombination
occurs. This can be done by creating a small cold spot
having a lower temperature than the rest of the cell walls.  The atoms
lost in recombination on the CS, are replenished from the bulk H
gas, which provides a large reservoir of
H\makebox[1.025ex][l]{$\downdownarrows$}- atoms and increases the
lifetime of the gas.

Adsorbed hydrogen atoms are in thermal equilibrium with the
ripplons of the  helium surface. This is ensured by the fast
momentum relaxation of adsorbed atoms caused by the
interactions with ripplons. At 100 mK the characteristic
relaxation time $\tau_{p}=3\times10^{-8}$ s is much shorter than
the surface residence time $\tau_{r}$ of hydrogen atom on liquid
$^4$He.\cite{Zimmerman1984} The recombination heat
accumulated into the system of H+ripplons is transferred to the sample
cell body through the phonons of the helium film and has to pass
through the ripplon-phonon thermal contact. The thermal coupling of H
to phonons of the film is rather weak.\cite{Walraven1990} The maximum
cooling 
power calculated for the phonon system at temperature $T_{ph}$ 
is given by\cite{RipplonPhonon}\begin{equation}
P_{rp}=G_{rp}(T_{s}^{20/3}-T_{ph}^{20/3}),\label{eq:RP
cooling}\end{equation} where $G_{rp}=0.84$
W$\textrm{cm}^{-2}\textrm{K}^{-20/3}$ is the heat conductance of
the ripplon-phonon contact. The cooling power is strongly
temperature dependent and $G_{rp}$ is smaller than the Kapitza
conductance across, e.g., metal/$^{4}\textrm{He}$ interfaces below
100 mK.\cite{Lounasmaa}

\section{Experimental}

Experiments were carried out in a top-loading cryostat with two
home-made refrigerators, a dilution refrigerator (DR) and a
$^{3}\textrm{He}$ refrigerator. The latter has been constructed to
cool a cryogenic rf dissociator, operating at 600-800 mK. The dissociator
contains a helical 350 MHz resonator and operates in a pulsed mode.
The construction of the dissociator is described elsewhere.
\cite{disso}  We were able to obtain a hydrogen flux reaching
the sample cell SC in the range 10$^{11}$-3$\times10^{13}$ atoms/s by
varying the pulse width (0.2-1 ms), the 
repetition rate (10-100 Hz) and the rf power fed into the
dissociator (1-10 mW).
The rf discharge also leads to
evaporation of the helium film in the dissociator, and the helium vapor
flows down and re-condenses in the filling line. To prevent the vapor
condensation in the sample cell we installed an accommodator into the
filling line just below the magnet fringe. The accommodator is
thermally anchored to the 0.3 K heat exchanger cooled by the
dilute stream entering into the still of the DR. After the
accommodator the H$\downarrow$
atoms pass through a hyperfine state polarizer, which is designed
to remove the $a$-state atoms. The polarizer is a cylindrical
chamber with several baffles inside to increase its surface area.
It is located in $\approx3.2$ T field and thermally anchored to
the lowest heat exchanger of the DR operating at a temperature range
from 100 to 120 mK. The temperature and magnetic field in the polarizer
are optimized to burn most of the $a$-state atoms before they
enter into the sample cell, and this way we obtain the maximum surface
density on the CS in the steady flux experiments as described below.

\begin{figure}
\begin{center}

\includegraphics[%
  width=0.80\linewidth,
  keepaspectratio]{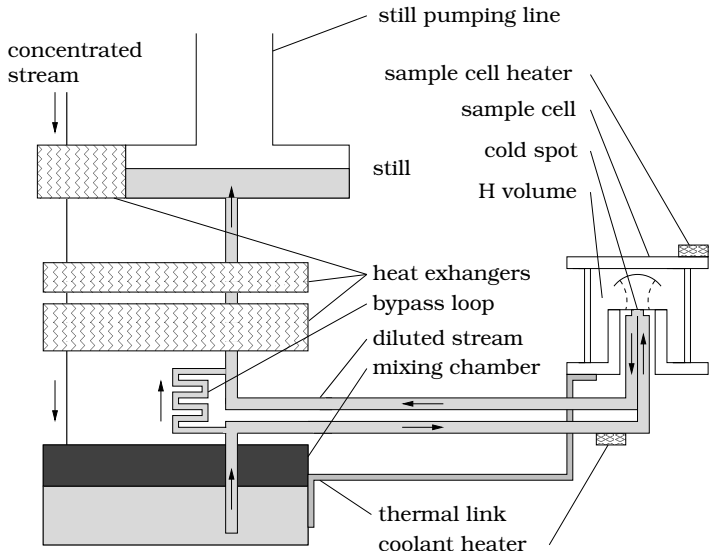}\end{center}

\caption{Cooling of the cold spot and the sample cell. The cold
spot temperature is set by the cooling liquid temperature. A
stable operation of the refrigerator is maintained by the by-pass
loop in the diluted stream.}
\label{cap:cryogenic}
\end{figure}

The sample cell is located in the center of a superconductive
magnet operating at 4.6 T and is thermally anchored to the mixing
chamber (MC) of the DR, which has the base temperature of 30 mK
and the cooling power of 150 $\mu$W at 100 mK. Due to a
relatively long distance of about 40 cm from the MC to the SC a
mechanical thermal anchor is not convenient for cooling a small
cold spot. Therefore, we utilized the dilute
$^{3}\textrm{He}-{}^{4}\textrm{He}$ ``coolant'' mixture from the mixing
chamber of the DR to cool the CS. Helium mixture is a
perfect dielectric and can fill up the mm-wave resonator
without disturbing its Q-value. At a typical circulation rate of
100 $\mu$moles/s the mixture flows with the speed of 2 cm/s in a
1.5 mm diameter tube and it has relatively large specific heat.
Another advantage of this method is a possibility of rapid (2-3
sec) change of the coolant temperature ($T_l$) with a heater
attached to the coolant tube going to the cell. If the whole
dilute stream of the DR would be used to cool the cold spot, then, at
high $T_{l}$, the operation of the DR would be seriously disturbed.
Therefore, the stream is divided into two approximately equal
parts as illustrated in fig. \ref{cap:cryogenic}.
Half of the stream flows through a by-pass loop and merges the CS
stream before entering into the DR heat exchanger. With this
construction we can raise $T_{l}$ up to 300 mK while still being
able to keep the SC and MC temperatures below 50 mK. Temperatures
of the mixing chamber, the sample cell ($T_{c}$), and the cold
spot coolant are measured with RuO$_{2}$ chip resistors
calibrated against the $^{3}$He melting curve thermometer and the NBS
SRM 768 superconducting fixed-point device. Estimated accuracy of
the thermometry is better than 1 mK.

\begin{figure}
\begin{center}

\includegraphics[%
  width=1.0\linewidth,
  keepaspectratio]{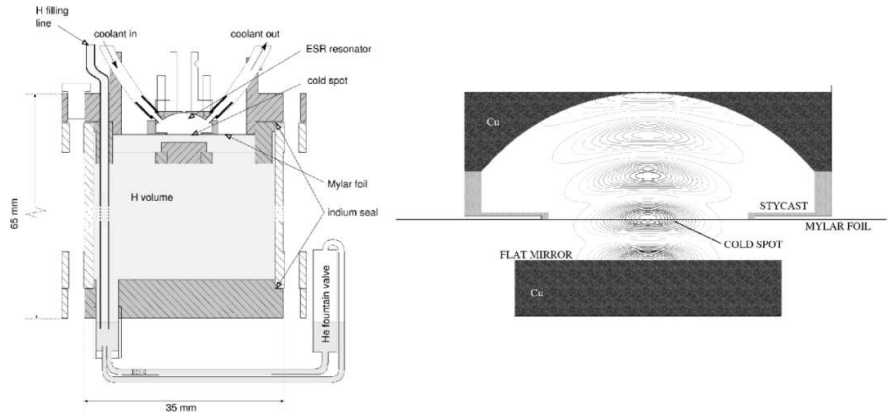}\end{center}

\caption{Left: The SCI with 6 mm diameter cold spot located on the lower
surface of the Mylar film forming the upper wall of the
SC. Right: The Fabry-Perot resonator and the cold spot in the SCI. The
contour plot shows the mm-wave field intensity profile of
TEM$_{003}$ mode.}
\label{cap:cell} 
\end{figure}

Electron-spin resonance (ESR) at 129 GHz is used to detect the
hydrogen atoms. The bulk and surface adsorbed atoms are in
contact with resonant mm-wave field in an open
Fabry-Perot resonator (FPR) similar to the optical resonators
used in lasers.\cite{kogelnik} Open geometry was chosen to ensure
an easy escape of the recombination products from the CS, which is
crucial in 2D compression experiments.\cite{MoskPHD} Schematic
drawings of the sample cell with the FPR and CS is shown in fig.
\ref{cap:cell}.
The upper mirror of the resonator is hemispherical with 6.5 mm
radius of curvature, while the lower mirror is flat. The spacing
between the mirrors is $\approx5.1$ mm corresponding to the
TEM$_{003}$ operating mode with loaded $Q\approx 1700$. The resonator
space is split in two parts,  isolating the SC from the
vacuum space, by a 20 $\mu$m Mylar film glued on
the top of the sample cell body. A Stycast 1266 ring, glued between
the semispherical 
mirror and the Mylar foil as shown in fig. \ref{cap:cell}, seals
the coolant volume. The coolant flows in and out of the resonator
through 1 mm i.d. copper tubes soldered into the semispherical
mirror.  The coolant stream flushes the upper surface of the Mylar
foil and thus
creates a 6 mm diameter cold spot on its lower surface. The
space in the resonator beneath the foil is filled by the hydrogen bulk
gas. This construction gives a good thermal isolation between the
CS and the SC, and makes it possible to use SC temperatures of
about 200 mK without a significant heat leak to the CS. Also,
the thermal boundary resistance between helium and Mylar is one of the
lowest of all 
solids, which provides the best cooling of the CS.  During operation
the FPR frequency is slightly dependent
on temperature. This is because, three quarters of the resonator
is filled with $^{3}$He/$^{4}$He mixture whose dielectric constant
is a function of the concentration ratio, which is temperature
dependent. Fortunately the dependence is rather weak at
$T_{l}\lesssim100$ mK and does not disturb noticeably the
stability of the ESR signal. On the other hand, by heating the
coolant to temperatures of the order of 1 K leads to an ESR
resonator frequency shift of about 50 MHz and helps to find the
resonance frequency after the cool down from room temperature.

The size of the gap between the cold spot and the lower FPR mirror
is very important because it influences the escape of the recombination energy
from the CS. In the first version of the sample cell described
above, the gap size was $\approx1.1$ mm, and from the measurements
(see Sec.4 and 5) we concluded that there is a significant
recombination overheating of the central part of the CS.
To reduce the overheating we have modified the FPR construction. The
lower mirror was replaced with hemispherical one, and moved
down by $\approx2.5$ mm increasing the gap to the CS by a
factor of $\sim3$ and changing the resonator mode to TEM$_{005}$.
As we demonstrate below, this led to a substantial improvement of
the heat removal from the CS. In this paper we present
experimental results obtained in both versions of the sample cell,
further on referred as SCI and SCII.

Another important parameter in the
thermal compression experiments is the size of the cold spot. In our
previous work\cite{Vasilyev2004,ESRinstab} we used 1.5 mm diameter cold
spot. This size has been chosen to minimize the total three-body
recombination rate and avoid overheating of the sample cell. It was
about twice smaller than the characteristic width of the mm-wave
profile (beam waist dia.$\approx3.2$ mm), which allowed imaging of
the density profile of the adsorbed gas near the edge of the
CS.\cite{Vasilyev2004} In the experiments we found that the
three-body recombination rate constant is $\approx20$ times
smaller than it was anticipated on the basis of existing data. The
contribution of the cold spot to the total loss rate was
negligibly small and we were not able to study recombination
processes on it. Therefore in the present setup we increased the
CS diameter to 6 mm. With this size most of the recombination occur on
the CS, and we were able to perform an accurate measurement of the three-body
recombination rate constant.\cite{Jarvinen2005} With the large spot
we lost the possibility to image the density distribution near the CS
edge, which went beyond the sensitivity region of the
FPR. On the other hand, probing the adsorbed gas at uniform
density offers a possibility to study the effects of
homogeneous ESR line broadening caused by interatomic collisions,
provided that other broadening mechanisms are reduced
to a negligible level.

Accurate measurements of the recombination heat released in the sample
requires that all the energy is
dissipated in the SC, and nothing is transferred back to the filling
line by the excited H$_2$ molecules. To prevent such a heat leak we
installed a $^{4}$He fountain valve into the H filling line at the
SC bottom as shown in fig. 2. The valve is controlled by a heater
and the helium level is measured with a capacitive level gauge.

The rate of the impurity-induced $b-a$ relaxation depends
on the quality of the sample cell walls.  To inhibit this unwanted
process we followed the standard procedure in the hydrogen
research. The SC was 
made of high-purity Cu and carefully etched in order to reduce the
relaxation. After that, we covered the inner surfaces of the cell
with polymer foil and epoxy. The impurity 
relaxation was further decreased by building up a layer of solid H$_{2}$ on
the SC walls. The layer is grown slowly during a week
when the H$\downarrow$ flux to the SC is kept on
continuously.\cite{StattH2layer} All these measures decreased the
intrinsic one-body relaxation rate to a negligible level
when compared to the three-body recombination rate on the CS.

The ESR spectrometer is based on a home-made 129 GHz heterodyne
receiver\cite{Vasiliev2004RSI} and its cryogenic part is thermally
anchored to the 1 K pot of the DR. The spectrometer is capable of
detecting both absorption and dispersion signals simultaneously
and has a sensitivity of about $10^{9}$ spins/G at mm-wave
excitation power of 20 pW.

\section{Measurement procedures and density calibrations}

Two types of measurements have been used to study
H$\downarrow$. In the decay type measurements the hydrogen sample is
first accumulated into the sample cell, the dissociator is turned
off and
the experimental conditions (temperature, volume or pressure) are
fixed. Then, the natural decay of the gas sample is
monitored. With this method most of the 
recombination mechanisms has been studied.\cite{bible} In the
second method, hydrogen source is running and filling the
sample cell continuously, so that a steady state is reached. The
equilibrium of the gas 
is balanced by incoming flux and recombination.\cite{ColdSpot,MoskPHD}
Main advantage of this method is a long term stability of bulk and
surface density and an easy control of recombination power by the
incident flux. In the present work most of the measurements were
of the decay type. The steady flux method was used only for
measuring the ripplon-phonon thermal contact as described at
the end of this section.

\begin{figure}
\includegraphics[width=0.9\linewidth]{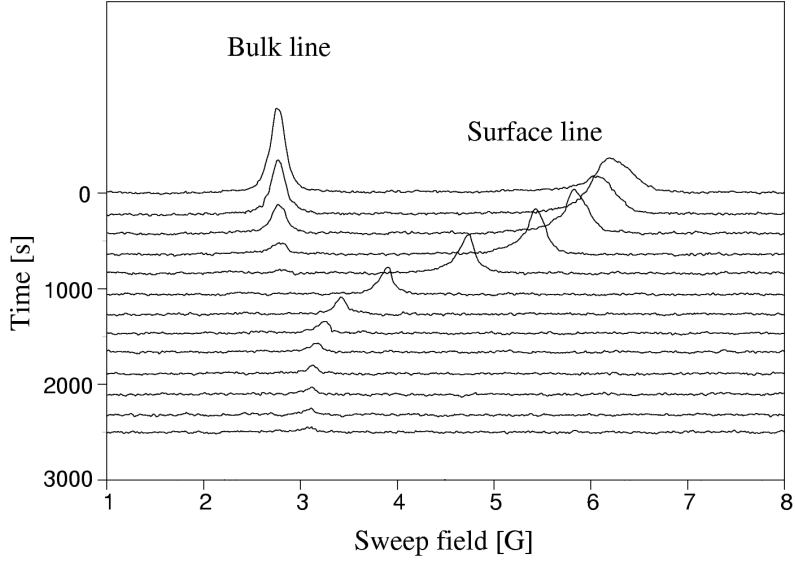}
\caption{Evolution of hydrogen ESR spectrum in a decay type measurement in SCI with $T_c=178$ mK and $T_l=78$ mK. The peak on the left is from the bulk gas and the peak on the right is from the 2D gas on the surface of the CS. The traces were recorded at intervals of 100 s.}
\label{cap:spektra}  
\end{figure}

\begin{figure}
\includegraphics[%
  clip,
  width=1.0\linewidth]{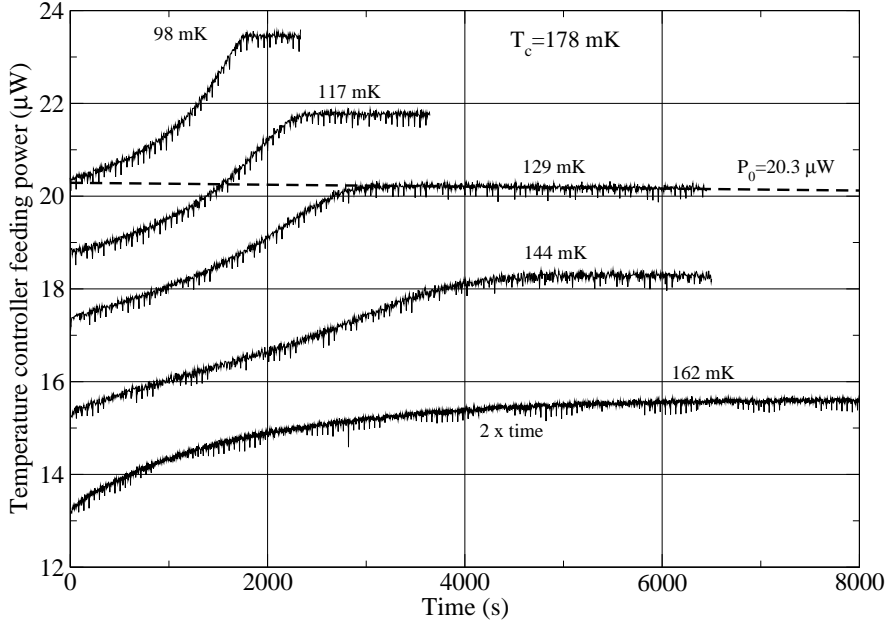}
\caption{The heating power $P(t)$ applied to sample cell by the
temperature controller at $T_{c}=178$ mK and $T_{l}$ ranging from
98 mK to 162 mK. The linear fit to the flat part of the signal
gives a background heating power $P_{0}=20.3$ $\mu$W for
$T_{l}=$129 mK. For the lowermost trace the time scale is
multiplied by 2.}
\label{cap:tcsignal}
\end{figure}

In the beginning of an experimental run a small amount (typically
1-4 mmoles) of H$_{2}$ is loaded into the dissociator from
room temperature. The H$_{2}$ gas freezes on the dissociator walls
and resonator helix. After that, we condense
$\approx8$ mmoles of isotopically purified $^{4}$He ($\lesssim1$ ppb of 
$^{3}$He) into the sample cell, required for proper operation of the 
fountain valve. Superfluid helium film is formed
lining the inner 
walls of the dissociator and the SC.  For the decay measurements
we run the dissociator at average power of 2-3 mW which is limited by the
back flow of helium into the dissociator along the filling line
walls. The highest flux obtained was 2-3$\times10^{13}$
atoms/s entering the sample cell. It usually takes about 2000 s to
reach the saturation bulk density of about 
$2\times10^{15}$ cm$^{-3}$. The hyperfine polarizer does not work
at high dissociator power, and the SC and MC are overheated to
$\approx250$ mK due to the $a-b$ recombination and re-condensing
helium vapors in the filling line and the SC. After switching off
the dissociator we cool down the mixing chamber, sample cell and
CS coolant to desired temperatures below 180 mK, which usually takes
5 to 10 min. During this time the $a$-state atoms are
burned in recombination and the
H\makebox[1.025ex][l]{$\downdownarrows$}-  sample is formed.
Typically, we start the decay measurements with the bulk density
of 1-5$\times10^{14}$ cm$^{-3}$ and we are able to reach the maximum
surface density of $5.5\times10^{12}$ cm$^{-2}$ on the CS. The
feedback power of the SC temperature controller and the ESR
spectra of bulk and surface atoms are the data collected during
the decay measurement. The raw data are presented in figs. \ref{cap:spektra} and \ref{cap:tcsignal}.

The ESR line of the adsorbed atoms is shifted from that
of the bulk gas due to the internal dipolar field created by the
polarized atoms in the plane where they are adsorbed. When the
plane is perpendicular to the polarizing magnetic field the dipolar
field is proportional to the surface density as $\Delta
H_{d}=-1.1(1)\times10^{-12}\sigma$
G$\cdot$cm$^{2}$.\cite{VasilievPRA2004,ShinkodaESR} Due to
rapid motion of the atoms, fluctuations of the dipolar field
are well averaged and the intrinsic linewidth of the surface gas
is expected to be very small. In practise the linewidth is determined by the
inhomogeneities of the polarizing magnetic field and/or surface
density inhomogeneities. The latter is especially important in the
thermal compression experiments due to an inhomogeneous
temperature profile near the cold spot edge.\cite{VasilievPRA2004}

The ESR line integrals of both the surface and bulk atoms are
proportional to the density of atoms. The absolute density
calibration is not an easy task, since we can not employ
calibration samples which are typically used for this
purpose in ESR spectroscopy. The calibration of the
H\makebox[1.025ex][l]{$\downdownarrows$}- bulk density is made
calorimetrically by measuring the recombination heat released in
the sample. We found this method more simple and reliable than the
calibration based on the absolute measurement of
microwave characteristics of the ESR resonator: coupling parameter
and Q-value.\cite{StattESR} The SC temperature controller adjusts
the heating power $P$ so that the sample cell temperature is
constant. A decrease or increase in the recombination power is
compensated by the temperature controller. In fig.
\ref{cap:tcsignal}
the recombination power $P$ is shown for several decays at
different temperatures of the CS cooling mixture. In these
experiments the number of atoms in the bulk of the sample cell is
much larger than on the surface. Then, by integrating the
recombination power, we calculate the bulk density
\begin{equation}
n(t)=\frac{2}{V_{h}D}\int_{t}^{\infty}[P_{0}-P(t')]dt',
\label{eq:Tem-density}
\end{equation}
where $V_{h}=40(1)$ cm$^3$ is the volume of the sample cell,
$D=4.478$ eV is the energy released in a single recombination
event,\cite{dissociation} and $P_{0}\equiv P(\infty)$ is the
stabilization power at $n=0$.

Similar procedure can not be used for measurement of the
surface density because the number of surface atoms in the ESR
resonator is negligibly small compared with that of the bulk gas.
It is known that the
strength of the bulk ESR signal is 
proportional to the product $n\cdot \int_{V_H} H_{1}^{2}dV$ with
$H_1$ being the magnetic component of mm-wave excitation and the
integral is taken over the volume occupied by hydrogen
gas.\cite{Poole1967} The surface line strength is, respectively,
proportional to $\sigma\cdot \int_{A} H_{1}^{2}dA$ where the
integral is taken over the area of the CS. To evaluate the ratio
of the integrals we solved numerically\cite{freefem} the mm-wave
field profile in the Fabry-Perot
resonator.
Once the ratio is known the surface density can be extracted by
using the calorimetrical calibration of the bulk density. Main errors
of the surface density measurement arise from
the inaccuracy of the mm-wave field calculation ($\sim5\%$) and
from the drifts of the ESR spectrometer excitation power and sensitivity
caused by the changes of liquid helium level in the cryostat. We
estimate that the overall error in $\sigma$ extracted from the ESR
integrals is about 10$\%$.

\begin{figure}
\includegraphics[%
width=1.0\linewidth]{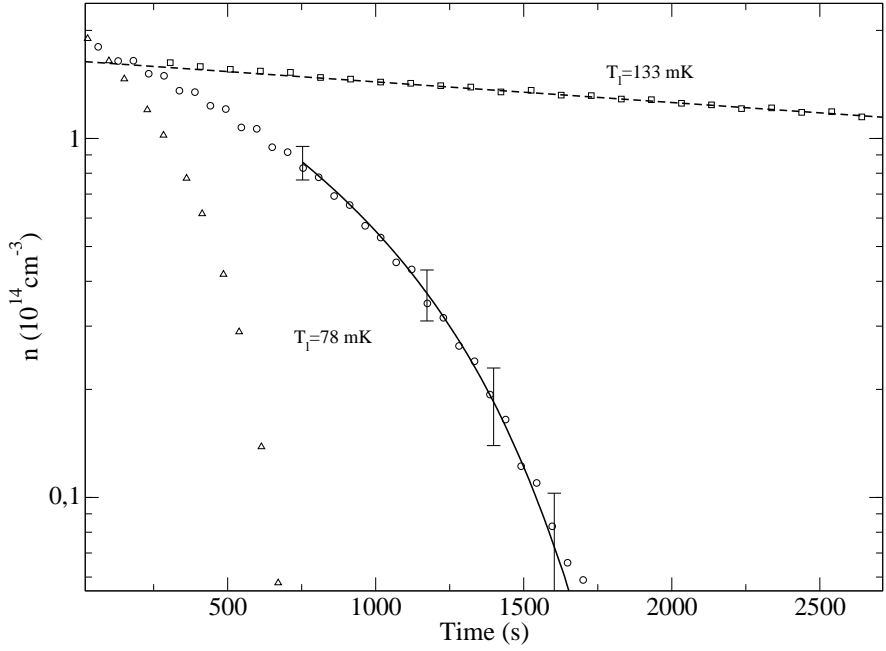} \caption{Bulk densities
extracted from the ESR spectra at $T_{c}=178$ mK.
$\square$-$T_{l}=133$ mK and $\circ-T_{l}=78$ mK with the SCI and
$\bigtriangleup$-$T_{l}=78$ mK with the SCII. 
The solid line with error bars is from a calorimetric
density calibration and the dashed line is an exponential fit to
ESR data points. The decays at $T_{l}=78$ mK are not exponentials
because of the high three-body recombination rate on the CS.}
\label{cap:Decay}
\end{figure}

Decays of H\makebox[1.025ex][l]{$\downdownarrows$}- in the
experimental conditions considered in this work are controlled by
two processes: one-body a-b relaxation and three-body
recombination on the walls of the SC and CS. They can be easily
distinguished by changing the CS temperature as demonstrated in
fig. \ref{cap:Decay}. At high enough temperature of the CS coolant
($T_l=133$ mK data in fig. \ref{cap:Decay}) surface density on the
CS is low and the influence of the three-body recombination is
negligibly small; thus the density decays exponentially as a
function of time. The slope of the decay curve in semi-log plot gives
$G_{1}^{e}$, the one-body relaxation rate. Cooling of the CS
strongly increases the 
loss rate ($T_l=78$ mK data in fig. \ref{cap:Decay}),
which is mainly caused by the three-body recombination on the
CS. In a general case the recombination rate can be found as
\begin{equation}
\frac{dn}{dt}=V_{h}G_{1}^{e}n+A_{s}L_{3}^{s}\sigma^{3},
\label{eq:recombination power}
\end{equation}
where $L_{3}^{s}$ is the surface three-body recombination rate
constant and $A_{s}$ is the area of the cold spot. By subtracting the
one-body contribution, extracted from the high $T_{l}$ data, it is
possible to obtain the value of the three-body recombination rate
constant $L_{3}^{s}=2.0(7)\times10^{-25}$ cm$^{4}$s$^{-1}$.\cite{Jarvinen2005}

In steady flux measurements the sample cell is continuously
filled with atoms. At the maximum dissociator power the hyperfine
polarizer is strongly overheated and the SC is being filled with atoms
in both $a$ and $b$ states in almost equal amounts. 
The total recombination power is so high that we can not get the
surface density on the CS higher than $10^{12}$ cm$^{-2}$. When the
dissociator power is decreased by about 20$\%$ we get the 
polarizer cold enough, and it effectively burns off most of the
\textit{a}-state atoms before they enter the SC. Then, a maximum
steady state surface density of
3.5$\times10^{12}$ is reached with a bulk
density of $n\sim10^{13}$ cm$^{-3}$. Further decrease of the
dissociator power leads to a decrease of the steady state
surface density. With the constant flux method we measure
steady state bulk and surface densities as functions of total
recombination power. Although we can not reach
high quantum degeneracy with this method, it allows 
to study the transfer of recombination heat through the helium
surface.

\section{Results and discussion}

\begin{figure}
\begin{center}
\includegraphics[%
  clip,
  width=1.0\linewidth]{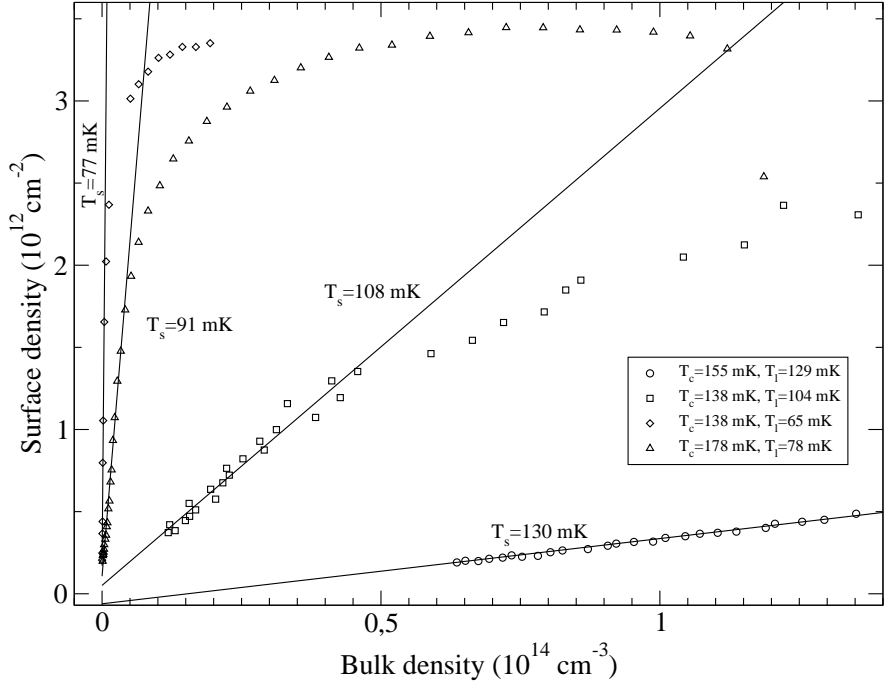}
\end{center}
\caption{Surface density plotted as a function of bulk density for
several decays at different $T_{c}$ and $T_{l}$. The surface
temperatures $T_{s}$ are extracted by fitting the adsorption
isotherm to the linear parts of the data (solid
lines).}
\label{cap:effective temp.}
\end{figure}

For an estimate of the quantum degeneracy reached in the experiments we
need to known the true temperature of the surface gas $T_{s}$. There is
no direct means to measure it, but it is possible to evaluate
$T_{s}$ from the adsorption isotherm eq. (\ref{eq:isotherm}) using the
known values of the bulk and surface gas densities. In fig.
\ref{cap:effective temp.} we present $\sigma$ as a function of $n$
measured at different coolant temperatures $T_{l}$. At high enough CS
temperature we are in the limit when the
recombination overheating is negligibly small, and  $T_{s}$
remains constant during the decay. Then, the surface density
increases linearly with $n$ according to eq. (\ref{eq:isotherm}).
This regime is seen in fig. \ref{cap:effective temp.} from the data
with $T_{l}=129$ mK. The slope of $\sigma(n)$ is proportional to
$\Lambda(T_{c}/T_{s})^{3/2}\exp(E_{a}/T_{s})$, and using the value
of $E_{a}/k_{B}=1.14(1)$ K we calculated the surface gas
temperature $T_{s}=130$ mK. This is, within the accuracy of our
thermometry, the same as the coolant temperature, which confirms the
reliability of the $T_s$ determination and the absence of overheating.
This measurement is the first experimental check of the adsorption
isotherm in the case of different bulk and surface gas
temperatures.
At lower coolant temperatures $\sigma(n)$ is not linear any more
and gets saturated at higher n. One can see in fig. \ref{cap:effective temp.}
that the saturation starts at lower n with the lower coolant temperatures.

First we consider only the low density region of the data, where $\sigma(n)$
is linear. The linearity of $\sigma(n)$ even for the
lowest $T_l$ implies
that the surface temperature $T_s$ from the
adsorption isotherm is constant. However, the difference between $T_s$
and $T_l$ 
is larger, the lower the temperature $T_l$ is. We consider two
possible explanations for this effect. (i) The CS surface is
overheated by the heat leak from the warmer CS walls along the
helium film. (ii) The 2D gas is not in equilibrium with the
bulk gas due to a hydrodynamic flow along the
surface.\cite{Safonov2005} 
The flow is driven by the gradient of 2D pressure
 at the edge of the CS. If the adsorption-desorption exchange
between the bulk and surface gas ($\sim n$) becomes slow enough, the
surface density  
on the CS may be substantially smaller than the equilibrium
value. Then $T_s$ extracted from
the adsorption isotherm is higher than the true surface
temperature. In the limit of ultra low density, when there is
practically no flux of atoms from the bulk, the 2D pressure $P\sim
n\cdot T$ is constant over the surface and thermal compression
cease to work. Measurements in this regime would give a nice
possibility to explore the equation of state of a 2D gas, which would
be especially interesting at quantum degeneracy. However, we found 
that an increase of the temperature difference between the SC and
the CS coolant (e.g. from $T_c-T_l=73$ mK to 100 mK in fig.
\ref{cap:effective temp.}) leads to an increase of the effective
surface temperature $T_s$. This confirms qualitatively that the CS
overheating at low $\sigma$ is caused by the heat flux from the
SC. Also the flow of adsorbed atoms out of the center of the cold
spot would create a nonuniform density profile with the density
decreasing from the center of the CS towards its edge. This would be
observed as a broadening of the left wing of the ESR lineshape, which
was seen previously on the smaller size cold
spot.\cite{VasilievPRA2004} In the present 
setup we have not observed such a broadening. Also, the analysis of the
hydrodynamic flow is complicated due to the gas of ripplons
on the helium surface.\cite{Safonov2005} The effects
of the flow are expected only at very small bulk and surface
densities, which are beyond the
sensitivity of our apparatus and thus making their experimental study
impossible at present. Therefore, in the following we concentrate
on the high density data, where the main goal of this work, high
quantum degeneracy, can be reached.

\begin{figure}
\begin{center}

\includegraphics[%
  clip,
  width=1.0\linewidth]{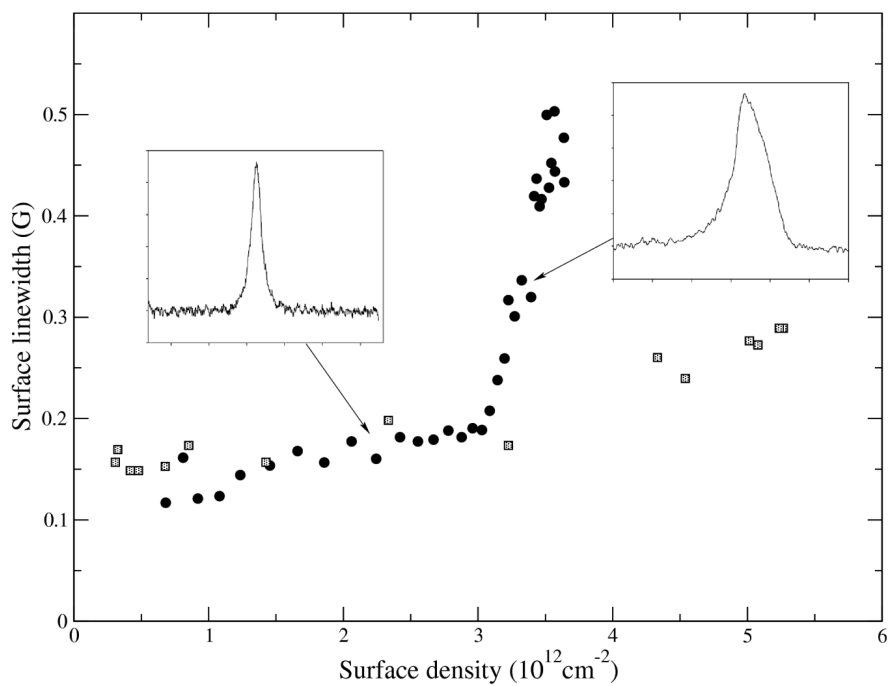}

\end{center}

\caption{Surface linewidth (FWHH) measured with SCI and SCII at similar conditions.
The insets shows two line shapes as examples demonstrating the appearance
of the asymmetry (same horizontal scale).}
\label{cap:Linewidth}
\end{figure}

At high surface density the 2D gas is overheated, which is revealed in the
leveling off of $\sigma$ with increasing $n$ in
fig. \ref{cap:effective temp.}. This is mainly caused by   
the three-body recombination.\cite{Jarvinen2005} 
At the same time in the SCI we
observed an abrupt increase of the surface ESR linewidth (FWHH),
as plotted in fig. \ref{cap:Linewidth}. The surface ESR line
changed from close to Lorentzian shape to asymmetrical shape with
strongly broadened right wing.
To understand the nature of the line broadening, we studied how
does the lineshape depend on the magnetic field gradients
$\frac{\partial B_{z}}{\partial r}$ created by a special shim coil
around the SC. By changing the direction and the magnitude of the
current in the coil we can compensate most of the inhomogeneities in
the internal dipolar field of the 2D gas. We found that the
broadening can be substantially reduced by the 
extra magnetic field with a parabolic radial profile and having the
minimum in the center of the CS. This means that the
internal dipolar field and the surface density profile have the
minimum in the center of the cold spot. This can be explained by
the presence of an extra heat source in the center of the CS, decreasing
towards the edge.

As already mentioned in sec. 2, it is very 
important in the compression experiments to ensure free escape of the
recombination products from the compression region. Although the
geometry of the Fabry-Perot  resonator was designed to be open, the
H$_{2}^{*}$ molecules 
formed on the CS are not completely free to leave the CS. The
flight path is restricted by the lower ESR mirror below the CS.
The molecules bounce from the lower mirror and may return back to
the CS. It is known that H$_{2}^{*}$ molecule needs approximately 150 surface
collisions with the helium film to release half of its internal
energy.\cite{vasiliev1993} The average H$_{2}^{*}$ excitation
energy released in a single wall collision is then about 170
K$\times k_{B}$. The presence of the lower mirror increases the
average number of collisions in the center of the CS compared that of
near the edge and may lead to the observed inhomogeneous 
overheating of the CS surface.

\begin{figure}
\begin{center}

\includegraphics[%
  width=1.0\linewidth]{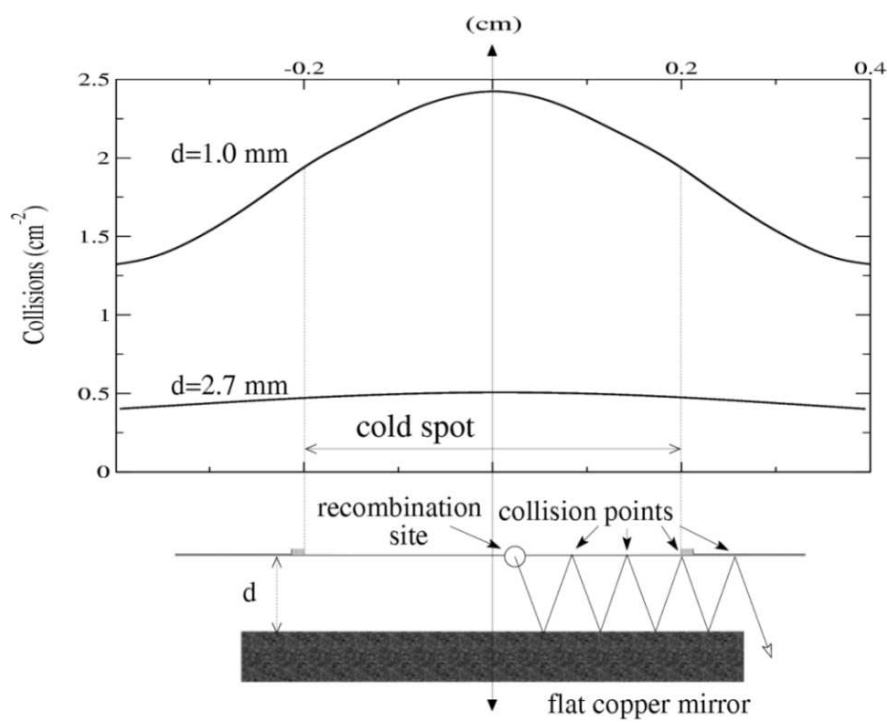}

\end{center}

\caption{Scheme of the CS overheating in SCII and the calculated flux of H$_{2}^{*}$
collisions per recombination event reflecting from the flat mirror.
The upper and the lower curve corresponds to different distances $d$ between
the flat mirror and the CS.}
\label{cap:Reflection heating}
\end{figure}

We performed a numerical simulation to estimate the extra heat
flux due to the reflecting molecules. In the simulation we assumed
that every molecule has a random recombination site on the CS and
a random direction of flight. The collisions with the CS and with the flat
mirror were considered to be elastic as illustrated in fig.
\ref{cap:Reflection heating}. After reaching the edge of the
mirror the molecule escapes and a new molecule is generated. We
calculated  $N_{col}$, the average number of collisions per unit area with the
CS surface per number of molecules generated, as a function of the
distance to the spot 
center (fig. \ref{cap:Reflection heating}). For the geometry of
the SCI with the flat ESR mirror being 1 mm below the CS, the
simulation gives parabolic H$_{2}^{*}$ flux profile with
$N_{col}=2.4$ cm$^{-2}$ in the center and decreasing by a factor
of 1.7 at the edge of the CS. This creates a heating of about
$120\textrm{ K}\times k_{B}\dot{\sigma}$ cm$^{-2}$, about half of
the recombination heat delivered directly to the surface when the
molecules are created. If the gap between the mirrors is increased
to 2.7 mm, the collision profile becomes almost flat and the number of
collisions is about $0.5$ cm$^{-2}$. This is still substantially
larger than that for a completely open cell where
$N_{col}=A_{c}^{-1}\approx0.01$ cm$^{-2}$ and $A_c$ is the area of the
SC. Once the reason for the
overheating was realized we modified the sample cell. The lower
mirror was moved down increasing the gap under the CS to 2.7 mm in
the SCII. We also had to
replace the flat mirror with a hemispherical one in order to have a
working resonance mode at so large spacing 
(TEM$_{006}$ mode). With these changes we eliminated the line
broadening effect, as demonstrated in fig. \ref{cap:Linewidth} and
succeeded to increase the maximum surface density from
3.6$\times10^{12}$ cm$^{-2}$ to 5.5$\times10^{12}$ cm$^{-2}$.

If the flat mirror in simulation is replaced with the spherical
mirror of SCII, the number of collisions on the CS increases by an
order of magnitude due to focusing effect of the spherical mirror.
This is in contradiction with the experiments with the SCII, where
the heating of the CS is clearly reduced. This means that the
collisions of the molecules with the helium film cannot be assumed
to be completely elastic. The inelastic collisions are not taken
into account in the simulation, which may be the reason why the model
fails to work for semispherical mirror.

\begin{figure}
\includegraphics[%
  clip,
  width=1.0\linewidth]{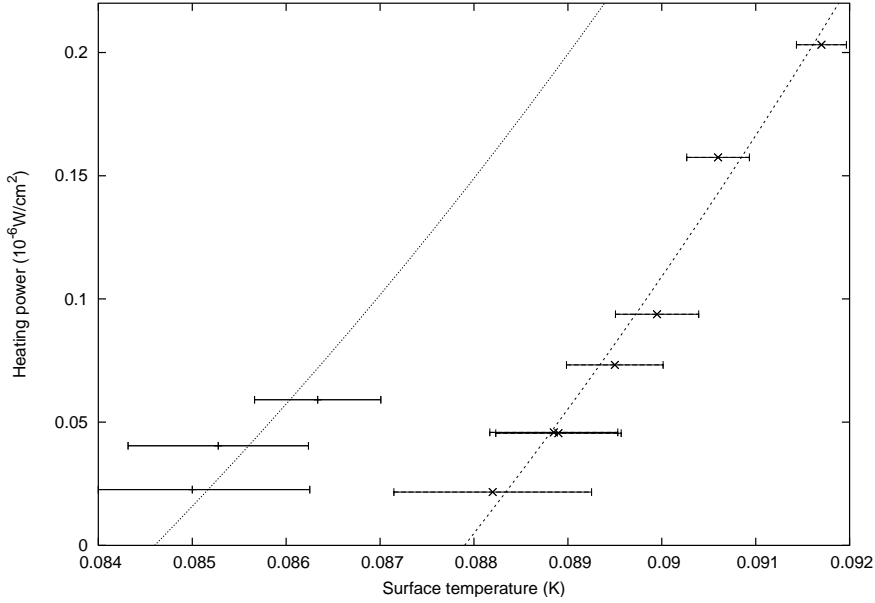}

\caption{Heating of the sample cell walls plotted as a function of $T_{s}$.
Different symbols corresponds to different coolant temperatures: $+-T_{l}=85$
mK and $\times-T_{l}=88$ mK. The lines are fits of the data to
eq. (\ref{eq:RP cooling}).}
\label{cap:RP contact}
\end{figure}

The thermal contact between the ripplons and phonons of helium
film $G_{rp}$ turns out to be a bottleneck of the recombination
energy removal from the cold spot.\cite{RipplonPhonon}  To
measure this quantity one needs to know the heat flux and both the
phonon and ripplon temperatures. In this experiment we used the
constant flux method. By measuring the bulk and surface densities we
extracted the ripplon temperature $T_s$ as described above. There
is no means to measure the temperature of the phonons of the
helium film, but since the thermal contact between the helium film
and the coolant is much better than the ripplon-phonon contact,
the phonon temperature is close to $T_{l}$. Furthermore, if
$T_l \ll T_s$, the $T_l^{20/3}$ term in eq. (\ref{eq:RP cooling})
is negligibly small making the accurate determination of $T_l$
unnecessary. In the constant flux method, all the atoms entering
the sample cell will recombine there in the exchange \textit{a-b}
recombination, thus, the total loss rate is equal to the input flux and
can be adjusted by the dissociator power. However, to know the
heat flux through the CS surface, we have to make sure that the
recombination energy is evenly distributed over the cell walls and the
CS. This is not the case if the CS temperature is much lower
than the cell temperature. Therefore, in the $G_{rp}$ measurements,
we kept the CS and SC temperatures rather close to each other,
so that the rate of the \textit{a-b} recombination on the cell
walls is much larger than that on the cold spot, i.e. $A_c
\exp(2E_a/T_c)/A_s\exp(2E_a/T_s)\gg1$ with $A_c$ and $A_s$ being,
respectively, the 
areas of the SC and the CS. Then the share of the total
recombination power absorbed by the CS is equal to $P_{rp}=P\cdot
A_s/A_c$, with $P$ being the total recombination power measured by
the cell temperature controller. In fig. \ref{cap:RP contact} we
plot the heat flux trough the unit area versus the surface
temperature $T_{s}$ for two different coolant temperatures.
Clearly the cooling power of the CS is strongly dependent on the
difference between $T_{s}$ and $T_{l}$.
The data of fig. \ref{cap:RP contact} were fitted to eq.
(\ref{eq:RP cooling}) with $G_{rp}$ and $T_{l}$ as free
parameters. The fit gives $G_{rp}=7\pm2$ W/cm$^{2}$ and $T_{l}$
values deviating less then 1 mK from the thermometer reading. The
result for $G_{rp}$ is about 8 times larger than the theoretical
value.\cite{RipplonPhonon} The discrepancy may arise from our
assumption that all the recombination heat is dumped to ripplons.
A large fraction can be transferred directly to phonons or rotons
of the helium film via the recondensing helium atoms. Therefore
our result can be considered as an upper limit estimate of $G_{rp}$. To our
knowledge, this is the first measurement of the ripplon-phonon
thermal conductivity.

\section{Limitations and future prospects}

\begin{figure}
\begin{center}

\includegraphics[%
  width=1.0\linewidth]{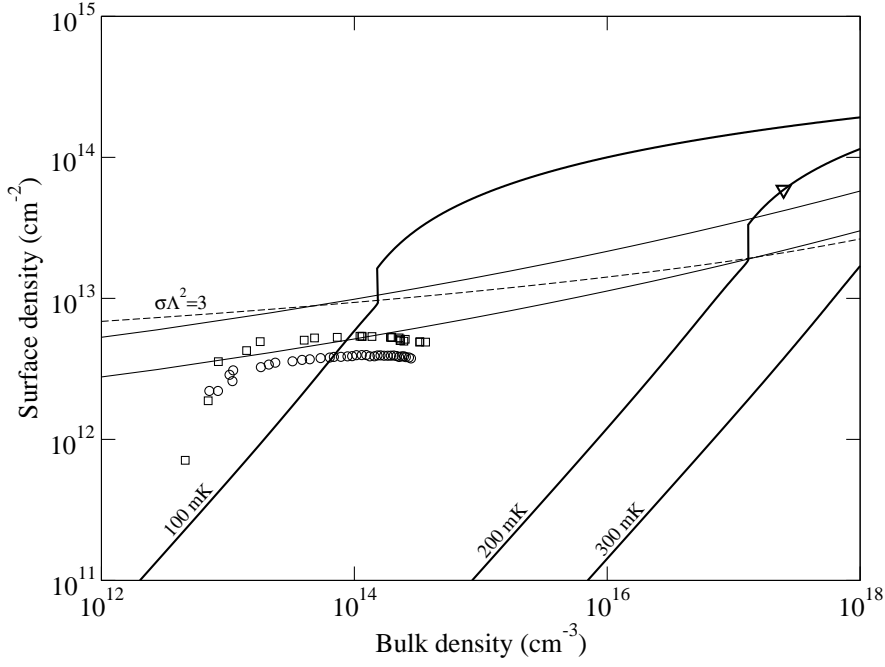}

\end{center}

\caption{Thick solid lines: Calculated isotherms of the surface gas. Dashed
line: Quasi-condensate formation at $\sigma\Lambda^{2}=3$. Thin solid
lines: Values calculated from eq. (\ref{eq:varpi heat balance}).
The lower line is calculated by using the theoretical value of $G_{rp}$
and the upper line by using the experimental value obtained in this
work. Experimental data ($\circ$) and ($\square$) are measured at
$T_{l}=67$ mK and $T_{c}=178$ mK respectively with SCI and SCII.
The point ($\bigtriangledown$) corresponds to the maximum 2D quantum
degeneracy achieved with magnetic compression.\cite{TurkuBKT}}
\label{cap:2D isotherms}
\end{figure}

In this work the highest surface density observed was about
$5.5\times10^{12}\textrm{ cm}^{-2}$ 
at $T_{s}\approx110$ mK, which corresponds to the 2D phase-space density
$\sigma\Lambda^{2}\approx1.5$. The dense surface gas was overheated
well above the coolant temperature and $\sigma$ was almost independent
of the bulk gas density as shown in fig. \ref{cap:effective temp.}.
The main source of heat in the SCII was the heat transferred to the
CS surface directly from surface recombination. Taking
into account 
only the heat of direct three-body 
surface recombination, a simple equation for the maximum quantum degeneracy
$\sigma\Lambda^{2}$ on the CS can be derived. The recombination heat
absorbed by the surface of the cold spot is given by

\begin{equation}
P_{3b}=(f+(1-f)\frac{A_{s}}{A_{c}})\frac{D}{2}L_{3}^{s}\sigma^{3},\label{eq:Direct three body heating}\end{equation}
where $f$ is the fraction
of the heat dumped directly to the cold spot surface $(f\leq2$ \%
\cite{ColdSpot,vasiliev1993,Meyer1994}). The efficiency with which
this heat can be removed is defined by the ripplon-phonon thermal
contact (eq. (\ref{eq:RP cooling})). Setting $P_{3b}=P_{rp}$ we
get an estimate of the highest achievable quantum degeneracy as

\begin{equation}
\varpi_{2}\equiv\sigma\Lambda^{2}=\frac{2\pi\hbar^{2}}{mk_{B}}\sqrt[3]{\frac{2G_{rp}}{DL_{3}^{s}(f+(1-f)\frac{A_{s}}{A_{c}})}}T_{s}^{\frac{11}{9}}\approx44T_{s}^{\frac{11}{9}},\label{eq:varpi
heat balance}\end{equation}
where $G_{rp}=7$ W$\textrm{cm}^{-2}\textrm{K}^{-20/3}$ has been
used to get the last form. In fig. \ref{cap:2D isotherms} we
present the results of thermal compression experiments with the
highest quantum degeneracy together with 
the plots of the adsorption isotherm at several surface temperatures.
Two sets of data are plotted, one with SCI and the other with
SCII. The result of the improvement can be clearly seen as a
factor of 1.5 increase in the surface density. The dashed line
marks the condition of quasi-condensate formation $\varpi_{2}=3$.
\cite{TurkuBKT} We include an abrupt change of the interactions
in 2D gas, which leads to an increase of $\sigma$ by a factor of
two in the quasi-condensate.\cite{Svistunov91} Eq. (\ref{eq:varpi
heat balance}) together with the adsorption isotherm
(\ref{eq:isotherm}) defines the maximum surface
density $\sigma(n)$, which can be reached for a given bulk density
$n$ and temperature $T_{c}$. For $T_{c}=178$ mK we plot
$\sigma(n)$ in fig. \ref{cap:2D isotherms} as the thin solid
lines. The lower line corresponds to the theoretical value of
$G_{rp}$ and the upper line is calculated using the experimental
value obtained in previous chapter. The SCII data show that
$\sigma(n)$ is higher than it is possible according to the theoretical
$G_{rp}$ value, but the values are still well below the value
calculated with the
experimental $G_{rp}$. This can indicate that even in the SCII
the CS is overheated by the reflected molecules. Thus, by removing the
molecules, the density can be increased close to the
quasi-condensation limit. This can be done by replacing the lower
spherical mirror with a grid or mesh, allowing for most of the
excited molecules to fly through the mirror. The best solution would
be to make the cold spot 
area completely open. However, this is
difficult to combine with the detection of atoms using
a mm-wave resonator.

Another possibility to increase $\varpi_{2}$ according to eq.
(\ref{eq:RP cooling}) is to use higher $T_{s}$. As one can see in
fig. \ref{cap:2D isotherms}, the maximum achievable degeneracy
goes well above the KTT line at higher temperatures. This is
because the ripplon-phonon thermal conductance grows rapidly with
increasing temperature. However, the use of higher temperatures
at the same time requires a large increase of bulk density, which
would, e.g., at $T=200$ mK approach to 10$^{17}$ cm$^{-3}$.
At so high bulk densities one will come across with thermal
accommodation heating the surface of the CS. The heating power of
thermal accommodation is given by

\begin{equation}
P_{acc}=\frac{nvs}{4}2k_{B}(T_{c}-T_{s}),\label{eq:Accomodation
heating}\end{equation}
where  $v$ is the thermal velocity and $s$ is the sticking probability
of H atom to the $^4$He 
surface.\cite{berkhout1993,yu1993}
The heating $P_{acc}$ becomes comparable with the surface
three-body recombination heat flux if the temperature difference
exceeds 100 mK at $n=6\times10^{16}$ cm$^{-3}$. Therefore, a
further increase of $n$ does not help to reach a higher degeneracy
of the surface gas. The thermal accommodation heat can also
explain why the highest $n$ data in fig. \ref{cap:2D isotherms}
deviate from the prediction of eq. (\ref{eq:varpi heat balance}). 

The adsorption energy is one more parameter which one may try to
change in order to get a higher degeneracy. Increasing $E_a$
together with the temperature, so that 
the ratio $E_a/T_s$ remains constant, will allow to improve the
ripplon-phonon contact and keep the bulk density low enough. The
adsorption energy can be increased by utilizing a thin undersaturated 
helium film \cite{Godfried1985} at the CS. Working with the thinner
films may also have several other benefits. The rate constant of the
three-body recombination should decrease at higher $E_{a}$.
\cite{Jarvinen2005} Also, one may hope on some improvement of
the surface cooling due to a direct interaction of the ripplons with
the phonons of the solid substrate. Making a superfluid helium
film locally thinner is, however, a difficult cryogenic problem
with no ready solution known at present.
\begin{acknowledgements}
We would like to thank  Michael
Hayden, Matti Krusius, Simo Jaakkola, and Kalle-Antti Suominen for
illustrative discussions. This work was supported by the Academy of
Finland and the Wihuri Foundation. J. J. acknowledge support from the
National Graduate School in Materials Physics. J. A. acknowledge
support from the National Graduate School in Nanoscience.

\end{acknowledgements}

\bibliographystyle{science}
\bibliography{bibloi}

\end{document}